\documentstyle[preprint,prl,aps]{revtex}
\begin{document}
%\twocolumn
\draft

\preprint{RU-94-94}
\title{ Chiral Non-Fermi Liquids in 1-d }
\author{Natan Andrei, Michael R. Douglas and  Andr\'es Jerez}
\address{Department of Physics and Astronomy, Rutgers University,
Piscataway, NJ 08855}
\maketitle

\begin{abstract}

We study models of chiral interacting fermions by means of conformal
 and Bethe-Ansatz techniques, and determine their thermodynamic properties
and asymptotic correlation functions.
We identify a class of  fixed points characterizing the infrared
behavior of the models. They display {\it chirally stabilized} non
 Fermi-Liquid behavior characterized by
universal exponents. A  realization of these
fluids may be found in the edge states of
 bilayered QHE systems. We calculate
the universal temperature dependence of the staggered conductance in
these systems.

\end{abstract}


%\narrowtext

Models displaying Non-Fermi-Liquid (NFL) behavior have been intensely
studied following the suggestion that the normal phase of
the  cuprate superconductors cannot be described in
 terms of the conventional Landau
theory \cite{pwa}.
The study of such models in $d \geq 2$ has made  slow progress. In
one dimension, however, a class of models was discovered, the Luttinger
Liquids, exhibiting NFL
behavior characterized  in the infrared (IR) limit
by single particle
correlation functions having only cuts with non-universal exponents.

In this letter we identify another large class of IR fixed points,
 available to
 interacting fermions with unequal number of left and right moving
degrees of freedom.
The new fixed points control the spin and (or) flavor sector of the
theory and
 are {\it universal} in character, independent of the strength
 of the interaction. We shall refer to
these fixed points as  chirally stabilized or coset fluids.

Several systems may be described by these chiral fluids. They must
break T-invariance, either by the presence of a magnetic field or by
the interactions.
Below we shall discuss in some detail one example,
the bilayered Quantum Hall system,
and present  predictions for the temperature dependence of its
staggered conductance.

The models we study tend in the ultraviolet (UV) to the free
hamiltonian, \\
$H_0=-i v_F \int dx \left(\sum _{r=1}^{f_R}\psi_{R,a,r}^*(x)\partial_x
\psi_{R,a,r}(x) -
\sum_{l=1}^{f_L}\psi_{L,a,l}^*(x)\partial_x \psi_{L,a,l}(x)\right)
$.  The fields $\psi^*_{R,a,r}(x)$ ($\psi^*_{L,a,l}(x)$),  with
 $a=\pm 1$  the spin
index, and   $r=1...f_{R},(l=1...f_{L})$ the right (left) flavor index,
 create
right-(left-) moving particles with a linearized dispersion
$\epsilon = \pm v_F (k-k_F)$.
 These particles are conventionally considered in the
Fock basis generated by the Fourier modes of $\psi^*_{R,a,r}(x)$ and $
\psi^*_{L,a,l}(x)$. As such $H_0$
 is an example of a (non-interacting) Fermi liquid.

The kinematics of one dimension permits many other bases to describe
particles with linearize dispersion, since
a linear combination of any number of  left (right)-movers is again a
left (right)-mover. This also
manifests itself in the operator language, in the
representation of
the field via abelian (or non-abelian) bosonization allowing
 a  separation of the charge, flavor and spin components.
The abelian basis $U(1)^{f_L}\otimes
U(1)^{f_R}$
is  convenient when the interaction terms involve the
charge densities,
$
\rho_{L}(x)=\sum_{l}\rho_{L,l}(x)\equiv
 \sum_{a,l}\psi^*_{L,a,l}(x)\psi_{L,a,l}(x)
$ (similarly for $\rho_R$). In this
 basis the fermionic fields take the form,
$
\psi_{L,a,l}=e^{i\phi_{L,a,l}(x)}$ with $\phi_{L,a,l}(x)$ a left
moving bosonic field.
The non-abelian basis $
\left(SU(2)_{f_L}\times SU(f_L)_2\times U(1)\right)\otimes
\left(SU(2)_{f_R}\times SU(f_R)_2\times U(1)\right)
$
is convenient when the interactions are mediated by the spin-densities
$S_L^i, S_R^i $,
where $S^i_L(x)=\sum_l S^i_{L,l}(x)\equiv \sum_{a,l}\psi^*_{L,a,l}
\sigma^i_{a,b} \psi_{L,b,l}
$ (similarly for $S^i_R$),
 which close on   Kac-Moody algebras with
central charges $f_L$ and $f_R$, respectively.
 $H_0$ is then represented as a sum of
theories \cite{witten,MS} corresponding to the charge- spin- and
flavor- components  \cite{al},
and the  operators of the free fermi theory
are expressed in terms of the operators in the
component theories,
$
\psi_{R,a, r}(x) = g_R^a(x)~ h_R^r(x)~e^{i\phi_R(x)}
$
where $g^a$ and $h^r$ are  fields in the spin and flavor
sectors, respectively, transforming in the
fundamental representation of
$SU(2)$ and $SU(f_R)$, and $\phi_R$ is the
charge field.  Each component of $H_0$ has its
 own  characteristics. In particular,
the spin component,
$
SU(2)_{f_L}\otimes SU(2)_{f_R}$, has the Virasoro charge
$c_L+c_R= 3f_L/(f_L+2)+3f_R/(f_R+2)$ and the
Kac-Moody charge $k_L+k_R=f_L+f_R$.
  Other bases are available: applying the Bethe Ansatz
 technique to the model allows the
construction of an arbitrary number of bases corresponding to a choice
of a scattering matrix between the left- (right-)
 movers $S_{LL} (S_{RR})$ \cite{nat}. Still the model
 describes a FL since the various components can be
put together to form a fermion.

When interactions are added the situation can change. To
lead to a  new behavior in the IR
the interaction needs to
 flow to some new fixed point, preferably at intermediate
coupling since strong coupling fixed points tend to open a gap.
The different components making up the electron are then sufficiently
modified, and one may find  that in the infrared the electron can no
longer be reconstituted. In other words, the natural degrees of
freedom will no longer be fermionic and one will observe a
NFL-behavior. This behavior is manifested in the structure
of the single particle correlation
functions at large distances, less so in the higher correlation functions and
 the low temperature thermodynamics. Thus  the specific heat  of any model
characterized in the IR by some conformal fixed point hamiltonian will be
will be linear, $c_V= {\pi \over 12}(c_L+c_R) T$, and the magnetic
susceptibility is a constant
$\chi=(k_L+k_R) \nu_0$,
($\nu_0= 1/ \pi v_F $ the density of states per unit length)
whether the fixed point describes a FL or not.

A familiar example is
the Luttinger Model, obtained by modifying the charge sector. The
hamiltonian is (here we choose $f_R=f_L$),
$
H=H_0 + g_{2c} \int dx \rho_L(x)\rho_R(x)+
g_{4c} \int dx \left( \rho_L(x)\rho_L(x)+ \rho_R(x)\rho_R(x)\right).
$
Renormalization group calculations \cite{sol} indicate that
 the model is conformally invariant (in the
infinite cut-off limit.)
As is well known, upon expressing the hamiltonian in terms of bosonic
fields it becomes quadratic and can be solved by a Bogoliubov rotation.
 The $g_4$ term modifies the spin and charge velocities (without
destroying the FL property at the fermi surface \cite{voit}),
while the effect of the $g_2$ term is to modify the exponents of the
fermionic correlation functions destroying the pole structure
characteristic of a FL.
Many models - the  Luttinger liquids \cite{hald}-
exhibit this low-energy behavior: they flow in the infrared to a $c=1$
bosonic model characterized by continuous parameters with no fermionic
interpretation.

\bigskip

We proceed now to study models flowing in the IR to  fixed points
describing the class of universal non Fermi-liquids.
The models are obtained by
 adding spin-exchange interactions to $H_0$,
\begin{equation}
H=H_0+\sum_{r,l}  \int dx \; (g_s)^{rl}S^i_{R,r}(x) S^i_{L,l}(x)\; .
\label{eq}
\end{equation}
with $(g_s)$ a matrix of couplings. The models are isotropic when the
matrix is diagonal.

Standard calculations show that the perturbation destabilizes
 the weak coupling fixed point.
 To
 identify where the models flow to one may perform a strong coupling or a
large-$(f_L$+$f_R)$ expansion.
Instead we shall apply techniques from  conformal field theory and
 thermodynamic Bethe Ansatz (TBA) and conclude
that the models  flow to  fixed points, the {\it
chirally
stabilized } fluids, given by a particular class of conformal field
theories - the WZW coset models (see below).
 We shall show that they describe NFL behavior
by  studying the correlation
functions and the S-matrices.

Here we chose an interaction term modifying the spin sector only.
 Other interactions may be added. Terms of the form  $
g_c\int dx \rho_R(x) \rho_L(x) +  \int dx
\; (g_f)^{ab}F^{\lambda}_{R,a}(x)F^{\lambda}_{L,b}(x) $
($F^{\lambda}_{R,a}=\psi^*_{R,a,r}t^{\lambda}_{rr'}\psi_{R,a,r'}$ is
the flavor density) will
modify the charge and flavor sectors:
 the $g_c$ coupling will
turn the isolated  fixed point into a line of NFL-fixed points, while the
$g_f$ term will drive the model to a non-trivial fixed point in the
flavor sector. One may further
add terms of the form $\int \left( S^i_L(x) S^i_L(x) + S^i_R(x)
S^i_R(x)\right) $ or $
\int \left( F^{\lambda}_L(x)F^{\lambda}_L(x)
+F^{\lambda}_R(x)F^{\lambda}_R(x)\right) $
which will modify the spin and flavor velocities, respectively, or
also terms that break global spin (or flavor) invariance.
These issues will
be discussed elsewhere.

We begin by discussing the flavor-isotropic
model, characterized by one coupling $g_s$.
 The interaction breaks the  spin symmetry to a global $SU(2)$. Therefore
 it is no longer conformally invariant, its IR
properties will depend on $f_R$--$f_L$.
For $f_R$=$f_L$, the model is chirally invariant
 and  flows to a strong coupling fixed point generating a
mass gap.
For $f_R>f_L$, on the other hand, the model is gapless and flows
 to a non-trivial fixed point which we proceed to identify.
This is possible since in this case the model is chiral in a strong
sense:
the conformal central
charges   as well as the Kac-Moody central charges  on left and
right are
 different.  The  charge differences $c_R$-$c_L$ and
 $f_R$-$f_L$ must, however, be preserved under
 the flow! \cite{foot}. This places
a strong constraint on the IR fixed point  - there is a unique
fixed point theory of lowest central charge satisfying these
two conditions. Thus,

%\widetext

\begin{equation}
SU(2)_{f_L}\otimes SU(2)_{f_R}\; \longrightarrow \;{SU(2)_{f_L}\times
 SU(2)_{f_R-f_L}\over SU(2)_{f_R}} \otimes SU(2)_{f_R-f_L}. \nonumber
\end{equation}

%\narrowtext

Here the Kac-Moody central charge $f_R-f_L$
 is matched by postulating
that the right degrees of freedom are those of the chiral WZW model
$SU(2)_{f_R-f_L}$ while the left degrees of
 freedom are singlet in spin and are described by
chiral coset CFT \cite{cft}.

The specific heat and the magnetic susceptibility can be immediately
 determined  from the IR central charges of the
 Virasoro and Kac-Moody algebras in the IR theory.
Hence the specific heat will be linear in the temperature
in the UV and in the IR
 limits  (with corrections \cite{corr}), and will undergo the flow:
\begin{eqnarray}
c_V^{uv}={\pi \over 6} ( f_L+f_R)T\; \longrightarrow\;
c_V^{ir}={\pi \over 6} \left(f_L+f_R +{3(f_R-f_L) \over f_R-f_L +2}
-{3f_R \over f_R +2} \right)T  \nonumber
\end{eqnarray}
where we also included charge and flavor contributions.
The flow in the suceptibility will be,
 $\chi^{uv}=(f_R+f_L) \nu_0 \rightarrow \chi^{ir}= (f_R-f_L) \nu_0$,
leading to a Wilson ratio $R_W=\left(f_L+f_R +{3(f_R-f_L)
\over f_R-f_L +2}
-{3f_R \over f_R +2} \right)/(f_R-f_L)$. To determine
the scales where the crossover in behavior occurs one needs to construct
the complete theory, this is done below through a Bethe-Ansatz.

We discuss now the operators around the IR fixed point. A $SU(2)_k$ theory
contains primary fields $\phi^{jm,\bar{j}\bar{m}}(x,t)$
transforming under a particular left and right representation of the
symmetry. There is a finite number of operators allowed: $0 \le
j,\bar{j} \le k/2$, and their dimension is $h = {j(j+1) \over k+2}$.
For the coset theory   there is a single primary
$\phi^{j,j'}_{j''}$ for each choice $0\le j\le f_L/2$,
$0\le j'\le (f_R-f_L)/2$ and $0\le j''\le f_R/2$.
The dimension of the primary is the difference of the dimensions of
the group primaries, up to an integer. We can thus match the physical
fields with the operator basis around the fixed point and  read off
the IR behaviour of the correlation functions,
\begin{eqnarray}
<\psi^*_{L,a,l}(x,t)\psi_{L,a',l'}(0,0)>
&\rightarrow&  \delta^{a a'}
\delta^{l l'}
(x-v_Ft)^{-(1+\delta_L )}
(x+v_Ft)^{-\delta_L } \nonumber \\
<\psi^*_{R,a,r}(x,t)\psi_{R,a',r'}(0,0)>&\rightarrow&
\delta^{a a'}
\delta^{rr'} (x-v_Ft)^{-\delta_R }(x+v_Ft)^{-(1+\delta_R)}  \nonumber \\
< S^i_L(x, t) S^j_L(0, 0)> &\rightarrow&
\delta^{ij} (x+v_Ft)^{-2}
 (x^2-v^2_Ft^2)^{-4/(f_R+2)}
\nonumber \\
< S^i_R(x, t) S^j_R(0,0)> &\rightarrow& \delta^{ij}
 (x-v_Ft)^{-2}
\nonumber
\end{eqnarray}
with $\delta_L=3/2(f_R+2)$ and $\delta_R=3f_L
/2(f_R-f_L+2)(f_R-f_L)$. We observe that the FL
 structure is destroyed: the momentum
distributions for small momenta are $ n_{\alpha}(k)
  \sim |k-k_F|^{2\delta_\alpha}$.
In these expressions the left and right components move with velocity
$v_F$. The inclusion of the term
$\int \left(J^i_LJ^i_L+J^i_RJ^i_R \right)$ would modify this. Also,
the charge and flavor correlation function will be non-trivial upon
inclusion of the terms mentioned earlier.

We consider now the case with flavor anisotropy.
Begin by studying the various
limits of extreme anisotropy, which can be modeled as a sequence of flows,
each
of the type described above.  Consider for example a coupling
$g^{rl}=g_1$ for $r\le f_{R1}$ and $g^{rl}=g_2$ for $r>f_{R1}$.
This breaks the $SU(f_R)\times U(1)$ right flavor and charge symmetry
down to
$SU(f_{R1})\times U(1)\times SU(f_{R2})\times U(1)$ with
$f_{R1}+f_{R2}=f_R$.
Clearly we want to bosonize the two groups of right fermions separately,
introducing spin densities $S^i_{R1}$ and $S^i_{R2}$ generating $SU(2)$
Kac-Moody algebras of level $f_{R1}$ and $f_{R2}$, so that
the interaction will again involve only the spin sector of the theory.

In the limit $g_1 \gg g_2$, we can regard the $g_1$ interaction as
generating
precisely the flow described above, approaching arbitrarily closely to
the IR
fixed point described above.  We can then identify the $g_2$ interaction
as a
specific perturbation of this IR fixed point using our earlier results.
If it is still marginally relevant, this will produce a flow to a
final IR fixed point.
Of course this analysis would be reversed for $g_2 \ll g_1$.
For the intermediate regime, we can make a guess as to the likely behavior
by
appealing to the $c$-theorem: non-trivial flows in $1+1$ dimensions always
decrease $c$ \cite{Zamolodchikov}.
If we compare the two final IR fixed points reached by the two limits of
extreme anisotropy and find that one has higher $c$, it is likely that any
finite anisotropy will cause the flow to continue to the other IR fixed
point.

There are several patterns which can arise in our example.
If $f_{R1}\ge f_L\ge f_{R2}$, the $g_1 \gg g_2$ limit will start with the
flow $
SU(2)_{f_L}\otimes SU(2)_{f_{R1}}\times SU(2)_{f_{R2}} \rightarrow
{SU(2)_{f_L}\times SU(2)_{f_{R1}-L}\over SU(2)_{f_{R1}}} \otimes
SU(2)_{f_{R1}-{f_L}}\times SU(2)_{f_{R2}}. $
The remaining interaction is irrelevant at this fixed point.
The $g_2 \gg g_1$ limit will follow a different  sequence:
$
SU(2)_{f_L}\otimes SU(2)_{f_{R2}}\times SU(2)_{f_{R1}}
\rightarrow
SU(2)_{{f_L}-f_{R2}}\otimes
{SU(2)_{f_{R2}}\times SU(2)_{{f_L}-f_{R2}}\over SU(2)_{f_L}}
\times SU(2)_{f_{R1}} \rightarrow
{SU(2)_{{f_L}-f_{R2}}\times SU(2)_{f_{R1}+f_{R2}-{f_L}}\over
SU(2)_{f_{R1}}}\otimes
{SU(2)_{f_{R2}}\times SU(2)_{{f_L}-f_{R2}}\over SU(2)_{f_L}}
\times SU(2)_{f_{R1}+f_{R2}-{f_L}}.$
If $f_L\ge f_{R1}\ge f_{R2}$, the two limiting sequences are both of
the latter form
 -- precisely this if $g_2 \gg g_1$, and with $f_{R1}$ and
$f_{R2}$ interchanged for $g_1 \gg g_2$.
One can check that in either case, if $f_{R1}> f_{R2}$,
the
result of the $g_2 \gg g_1$ sequence always has lower $c$ than the
result of the
$g_1 \gg g_2$ sequence, making it the IR fixed point for generic
anisotropy. The correlation functions for this case can be
found by the same means and will be given in a subsequent work.

\bigskip

We can actually follow the flow at any scale by solving the model
exactly.
The model is closely related to
the multichannel Kondo model and exhibits dynamical fusion \cite{ad}
 (a different approach
was given in \cite{pw}) allowing a solution
  by a method very similar to the one used to solve the
anisotropic multichannel Kondo model \cite{pap}. We find that
the model generates scales $m^l_L,m^r_R: \; l\le f_L,~ r\le f_R$,
parametrizing the patterns of
flavor symmetry breaking, and setting the excitation energies and
momenta. The free energy is given by,

\begin{eqnarray}
F(T,h) =
-\frac{TL}{2\pi}\int_{-\infty}^{\infty} d\xi \left(\sum_r  m^r_R e^{-\xi}
\ln(1+\eta_{r}(\xi,\frac{h}{T})) + \sum_l  m^l_L e^\xi
\ln(1+\eta_{l}(\xi,\frac{h}{T}))\right), \nonumber
\end{eqnarray}
where the functions
$\{\eta(\xi,\frac{h}{T})\}$ are the solution of the following
system of coupled integral equations (TBA-equations):
\begin{eqnarray}
\ln \eta_n =-2\frac{m^n_L}{T} e^\xi  -2\frac{m^n_R}{T}
e^{-\xi}  + G\ln(1+\eta_{n+1}) +
G\ln(1+\eta_{n-1}),~~~n=1,...,\infty,~~\eta_0\equiv 0, \nonumber
\end{eqnarray}
with boundary condition:
$\ln \eta_{n} \rightarrow 2n\mu h/T$. The integral operator
 $G$ is defined by the kernel
$1/(2\pi\cosh(\xi'-\xi))$. In the isotropic case, $m_L^l=m_L
\delta_{f_L,l}, \; m_R^r=m_R \delta_{f_R,r}$.
 The forms of the
driving terms $m_L e^\xi$ and $m_R e^{-\xi}$ are characteristic of
 massless left and
right moving excitations, and when both occur at the same level
 a driving term $m \cosh \xi$ results indicating a mass gap.
To be more explicit, consider the
case of  two-channels of  right movers and one-channel of left movers.
The two
 couplings $g_1,~g_2$, are obtained by diagonalizing
the matrix of couplings $g_s^{rl}$. Choose  $g_1 < g_2, \;\phi = g_1/g_2$.
The physical scales then are, $
m^1_R = 2D_R \cos (\frac{\pi \phi}{2}) e^{-\frac{\pi}{g_2}}; \; m_R^2 =
D_R e^{-\frac{\pi}{g_1}} \;$ and $
m_L^1 = 2D_L  e^{-\frac{\pi}{g_1}}
$, with $D_L, D_R$ the densities of left and right movers \cite{pap}.

 From the  TBA-equations both the IR and the UV limits can be read off using
TBA-rules \cite{ken}: the IR limit of the left movers is obtained from
 the equations by
considering the right-mass as infinitely heavy and
truncating the equations  at the level it was inserted. An analogous
rule holds
for the right movers. The UV limit is obtained, on the other hand, by
considering the masses as vanishing. Applying these rules we deduce
the IR limit, and find accord with the conformal considerations.
The solution of the equations provides in addition the full
interpolation between the UV and IR limits.

\bigskip

To observe  chiral NFL behavior in an experimental system one may study
 the edge states in a bilayer of a quantum Hall liquid.
In the higher hierarchy states (e.g. for filling factor $\nu=n/(np+1)$
with $p$ a negative even number) one edge state mode moves in a
direction opposite to the rest \cite{fisher} providing a chiral
imbalance. The
channel degrees of freedom play the role of the  flavor and the
bilayer provides a `spin' degree of freedom. The  spin-exchange
 interaction (eq.(\ref{eq})) will
be induced through virtual hoppings between the layers. We assume that
the
 couplings have been adjusted to have spin $SU(2)$ symmetry.
The  system
will then exhibit a universal behavior in the
staggered conductance  $G_s$ measuring the response to  electric
fields that are oppositely oriented in the layers. The staggered current is
given by $J_s= v_F(S^3_R-S^3_L)$, -- it is the spin current of the
model -- and  the
staggered current-current correlation function can be expressed in
terms of the spin density correlations calculated above. We
conclude  (setting $f_L=1, \; f_R=n-1$) that
$G_S \sim T^{8/(n+1) -1}$, and expect for $n=7$ a staggered
metal-insulator transition induced by correlations \cite{sp}.

\bigskip

Acknowledments: It is a pleasure to thank E. Y. Andrei, P. Coleman, K.
Intrilligator, L.
Ioffe,  A.
Ruckenstein and A. J. Schofield for comments, suggestions
and enlightening discussions.


\begin{references}
\bibitem{foot} A simple argument for this is that
a theory with $c_R\ne c_L$ has a gravitational anomaly which precludes
defining
it on a curved space (without introducing a local frame); if we add
additional
decoupled chiral degrees of freedom to cancel the anomaly, this
cancellation
must also work in the IR limit.
A similar argument can be made for the difference of the Kac-Moody central
charges $f_R-f_L$, by considering the anomaly in the vector spin current.
\bibitem{corr}  The leading
correction (in the infinit cut-off limit)
 to the linear term in the specific heat
 is $
 T^{2+4/(f_R-f_L+2)}$. It comes from the leading singlet
 irrelevant operator at the IR fixed point
$\bar\phi^{0,1}_{0} J^a g^{1a}$.
\bibitem{sp} In terms of the original variables we have a transition
in the spin-conductivity.
\end{references}
\end{document}